\documentclass{aa}
\usepackage{graphics}
\usepackage{times}
\begin{document}

\thesaurus{ 01    
              (02.08.1; 
        03.13.4; 
        08.19.4; 
                13.21.5)}

\title{UV Light Curves of Thermonuclear Supernovae}

\author{S.I. Blinnikov \inst{1,2,3}
   \and E.I. Sorokina \inst{4}}
\date{Received December 24, 1999 / Accepted March 6, 2000}

\institute{ Institute for Theoretical and Experimental Physics, 117218 Moscow,
Russia (sergei.blinnikov@itep.ru)
\and
Max-Planck-Institut f\"ur Astrophysik, D-85740 Garching, Germany
\and
Institute of  Laser Engineering, Osaka, 565-0871, Japan
\and
Sternberg Astronomical Institute, 119899 Moscow, Russia
(sorokina@sai.msu.su) }

\maketitle

\begin{abstract}

Ultraviolet light curves are calculated for several thermonuclear
supernova models using a multifrequency radiation hydrodynamic code.
It is found that Chan\-d\-ra\-se\-khar-mass models produce very similar light
curves both for detonation and deflagration.
Sub-Chandrasekhar-mass models essentially differ from ``normal''
Chandrasekhar ones regarding behaviour of their UV fluxes.
Differences in absolute brightness and in shape of light curves of
thermonuclear supernovae could be detectable up to 300~Mpc with modern UV
space telescopes.
\end{abstract}

\section{Introduction}

Early ultraviolet (UV) emission from Type Ia supernovae (SNe~Ia) is poorly
known by now.
Only a very few brightest events have been observed a decade ago
with the International Ultraviolet Explorer (IUE).
Although observational data were sometimes quite fascinating, as in the case
of SN 1990N (Leibundgut~et~al. \cite{leietal}, Jeffery~et~al. \cite{jefetal}),
an amount of observed UV light curves and spectra remained too small to reveal
what is typical for the UV emission from SNe~Ia and how individual features
of the explosion can be displayed.

The Hubble Space Telescope (HST) has slightly improved the situation.
More data of better quality were obtained.
This allowed theorists to make a comparison between the predictions
of explosion models and the observational results.
Reproduction of supernova UV emission is a good test for an explosion model
because this spectral region reflects more directly the distribution
of \element[][56]{Ni} synthesized during the explosion and the conditions
in the exploding star.
Several models were already used to fit the observed spectra.
Analyses of early and late emission from SNe~Ia
by Kirshner~et~al.~\cite{kiretal}, Ruiz-Lapuente~et~al.~\cite{rulaetal},
Nugent~et~al.~\cite{nugetal} show that Chandra\-se\-khar-mass models DD4
(Woosley \& Weaver~\cite{wwmods}) and W7 (Nomoto et al.~\cite{nthyo}) and
sub-Chandrasekhar-mass helium detonation models (see Livne~\cite{liv};
Livne \& Glasner~\cite{ligl}; Woosley \& Weaver~\cite{wwlowm};
H\"oflich \& Khokhlov~\cite{hoekho}) can reproduce some features of UV
spectra of SNe~Ia quite well.

In this Letter we calculate the light curves of the similar models and discuss
how they differ from each other in several UV wavelength ranges.
It is quite probable that more observational data will soon be available with
the HST, and that the Far-Ultraviolet Spectroscopic Explorer
(FUSE), operating at shorter wavelengths (Sembach~\cite{sem}), will be able
to obtain light curves and spectra of SNe~Ia in the range where they were not
observed so far.
The analysis proposed here can help to distinguish which mode of explosion is
actually realized in the SNe~Ia.

\section{Models and method of calculations}

In our analysis we have studied two Chandrasekhar-mass models: the classical
deflagration model W7
(Nomoto et al.~\cite{nthyo}) and the delayed detonation model DD4
(Woosley \& Weaver \cite{wwmods}), as well as two
sub-Chandrasekhar-mass models: helium detonation model 4
of Livne \& Arnett~(\cite{liar}) (hereafter, LA4) and low-mass detonation
model with low \element[][56]{Ni} production  (hereafter, WD065;
Ruiz-Lapuente~et~al.~\cite{rula91bg}).
Main parameters of these models are gathered in the Table~\ref{uvfluxes}, as
well as the values of rise time and maximum UV fluxes in
the standard IUE range (just as it is the most typical
form of representation; e.g., Pun~et~al.~\cite{punetal}) and in the FUSE
range. For definiteness the distance to a supernova is supposed to be 10~Mpc.

The method used here for light curve modeling is multi-energy group
radiation hydrodynamics.
Our code {\sc stella} (Blinnikov \& Bartunov~\cite{blba};
Blinnikov~et~al.~\cite{bebpw})
solves simultaneously hydrodynamic equations and time-dependent equations
for the angular
moments of
intensity averaged over fixed frequency bands, using up to $\sim200$ zones
for the Lagrangean coordinate and up to 100 frequency bins
(i.e., energy groups).
This allows us to have a reasonably accurate representation of
non-equilibrium continuum radiation in a self-consistent
calculation when no additional estimates of thermalization depth are needed.
Local Thermodynamic Equilibrium (LTE) for ionization and atomic level
populations is assumed in our modeling.
In the equation of state, LTE ionization and recombination are taken
into account.  The effect of line opacity is treated as an expansion
opacity according to the prescription of Eastman \& Pinto~(\cite{easpi})
and Blinnikov~(\cite{blinn96,blinn97}).

\section{Results and discussion}

\begin{figure}
\resizebox{\hsize}{\hsize}{\includegraphics{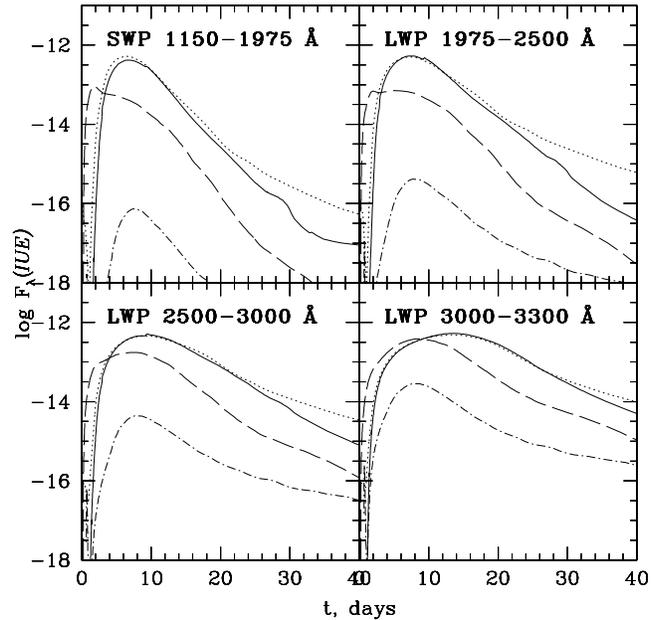}}
\caption{Near-UV light curves of the models DD4 (solid), W7 (dots), LA4
(long dash), and WD065 (dash-dot). Fluxes $F_\lambda$ in units
~ergs~s${}^{-1}$~cm${}^{-2}$~\AA${}^{-1}$ averaged
in the four IUE spectral bands are reduced to distance of 10~Mpc.}
\label{iue}
\end{figure}

\begin{figure}
\resizebox{\hsize}{\hsize}{\includegraphics{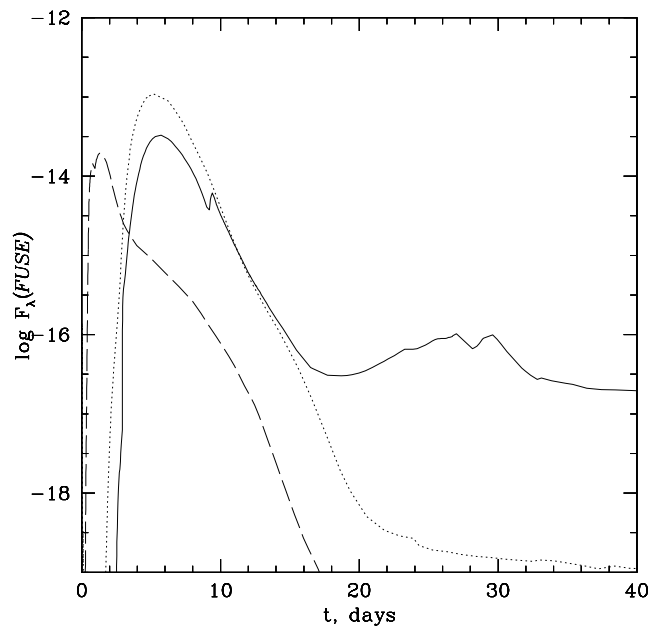}}
\caption{Far-UV light curves for the same models as in Fig.~\protect\ref{iue},
but in the FUSE working wavelength range (905--1187\AA)}
\label{fuse}
\end{figure}

The main results of our calculations are presented in
Figs.~\ref{iue},~\ref{fuse}.
The light curves of our models are shown in near and far UV ranges.
The fluxes are plotted as they would be seen for supernovae at distance of
10~Mpc.
Declared sensitivity of FUSE is
$\sim 3\cdot 10^{-15}$~ergs~s${}^{-1}$~cm${}^{-2}$~\AA${}^{-1}$, and sensitivity
of HST (at working range of wavelengths almost equal to that
of IUE) is roughly
$10^{-16}$~ergs~s${}^{-1}$~cm${}^{-2}$~\AA${}^{-1}$.
This allows us to estimate that SNe~Ia could be observed up to 300~Mpc in
the near UV and up to 30~Mpc in the far UV (with HST and FUSE, respectively).
Yet it should be noticed that in the far UV SNe~Ia are bright enough only
during several days, and their flux declines very quickly after the maximum
light, so the probability of discovering for them is quite low, unless they
are very close to us (a few Mpc or even less).

\begin{table}
\caption{Parameters of SN Ia models, rise time to the maximum of the UV light
curves and UV fluxes at maximum light in FUSE and IUE ranges.
Fluxes are calculated under the assumption that a supernova is
at distance of 10~Mpc from the observer.}
\label{uvfluxes}
\begin{tabular}{lllll}
\hline
\hline
Model & DD4 & W7 & LA4 & WD065 \\
\hline
$M_{\rm WD}{}^{\rm a}$      & 1.3861 & 1.3775 & 0.8678 & 0.6500 \\
$M_{{}^{56}{\rm Ni}}{}^{\rm a}$ & 0.63 & 0.60 & 0.47   & 0.05 \\
$E_{51}{}^{\rm b}$ & 1.23   & 1.20   & 1.15   & 0.56 \\
\hline

\multicolumn{5}{c}{FUSE 905--1187\AA \rule{0mm}{4mm}} \\
$t_{\rm max}{}^{\rm c}$ & 5.7 & 5.3 & 1.4 & 7.7 \\
$F_\lambda{}^{\rm d}$ 
                                        & 3.28
                                        & 11.0
                                        & 1.97
                                        & $4.83 \cdot 10^{-9}$ \\
\hline
\multicolumn{5}{c}{SWP 1150--1975\AA \rule{0mm}{4mm}} \\
$t_{\rm max}{}^{\rm c}$ & 6.7 & 6.4 & 1.8 & 7.7 \\
$F_\lambda{}^{\rm d}$ 
                                        & 41.7
                                        & 51.8
                                        & 9.21
                                        & $7.45 \cdot 10^{-3}$ \\
\hline
\multicolumn{5}{c}{LWPshort 1975--2500\AA \rule{0mm}{4mm}} \\
$t_{\rm max}{}^{\rm c}$ & 7.5 & 7.4 & 2.0 & 8.0 \\
$F_\lambda{}^{\rm d}$ 
                                        & 53.2
                                        & 50.4
                                        & 6.90
                                        & $4.22 \cdot 10^{-2}$ \\
\hline
\multicolumn{5}{c}{LWPmiddle 2500--3000\AA \rule{0mm}{4mm}} \\
$t_{\rm max}{}^{\rm c}$ & 8.7 & 9.1 & 7.5 & 8.1 \\
$F_\lambda{}^{\rm d}$ 
                                        & 46.0
                                        & 45.8
                                        & 17.7
                                        & 0.447 \\
\hline
\multicolumn{5}{c}{LWPlong 3000--3500\AA \rule{0mm}{4mm}} \\
$t_{\rm max}{}^{\rm c}$ & 13.5 & 13.4 & 8.5 & 8.2 \\
$F_\lambda{}^{\rm d}$ 
                                        & 52.7
                                        & 48.9
                                        & 38.3
                                        & 2.87 \\
\hline
\multicolumn{5}{c}{Total IUE luminosity 1150--3300\AA \rule{0mm}{4mm}} \\
$t_{\rm max}{}^{\rm c}$ & 7.9 & 7.4 & 7.6 & 8.2 \\
$L{}^{\rm e}$ 
                                        & 11.0
                                        & 11.4
                                        & 3.72
                                        & 0.161 \\
\hline
\multicolumn{5}{l}{${}^{\rm a}$in $M_{\sun}$} \\
\multicolumn{5}{l}{${}^{\rm b}$in $10^{51}$~ergs~s${}^{-1}$} \\
\multicolumn{5}{l}{${}^{\rm c}$in days} \\
\multicolumn{5}{l}%
{${}^{\rm d}$in $10^{-14}$~ergs~s${}^{-1}$~cm${}^{-2}$~\AA${}^{-1}$} \\
\multicolumn{5}{l}{${}^{\rm e}$in $10^{42}$~ergs~s${}^{-1}$} \\
\end{tabular}
\end{table}

Differences in the shapes and absolute brightness of SNe~Ia light curves
become more pronounced in this spectral range, especially in far UV,
for different modes of explosion.
As we have found in our paper (Sorokina~et~al.~\cite{sbb}),
the light curves of W7 and DD4 are very similar in {\it B} band close
to the maximum light and differ drastically several days after it.
We can see likely behaviour in near-UV wavelengths (Fig.~\ref{iue}).
In far UV, differences grow up, so that even at maximum light one can
distinguish between two Chandrasekhar-mass models.
Such a behaviour can most probably be explained by different distribution of
\element[][56]{Ni} inside debris of exploded star.
In our case, the fraction of \element[][56]{Ni} decreases more sharply
in the model W7, and this perhaps leads to the faster decline of its light
curve.

The shape of the UV light curve of WD065 is conformal to those of
Chandrasekhar-mass models, though it shows much lower absolute flux (due to
an order-of-magnitude lower \element[][56]{Ni} mass).
The light curve maxima of the Chandrasekhar-mass models are shifted
progressively towards later epochs for longer wavelengths remaining almost
equal in brightness, while the WD065
maxima occur at nearly the same epoch for all of the IUE and FUSE
ranges, and the emission virtually disappears at the shorter edge of the
spectrum (see Table~\ref{uvfluxes} and Fig.~\ref{iue}), since such a small
amount of \element[][56]{Ni} as present in this model is not able to maintain
high temperature inside the ejecta. The emission of WD065 in far UV
becomes so weak that it could be detected only if supernova exploded
in our neighbourhood (not farther than a few hundreds of parsecs from us).

The model LA4 (helium detonation in outer layers) is apparently
distinguished by its rise time in the shortest spectral bands.
The most interesting feature of this light curve is its clear two-maxima
structure in the short LWP range.
The earliest spike of the far-UV radiation is due to the outer
\element[][56]{Ni} layer specific for this model.
It is well known that those helium detonation models are too blue near
visual maximum (H\"oflich \& Khokhlov~\cite{hoekho};
Ruiz-Lapuente~et~al.~\cite{rulaproc}).
This is also confirmed by our {\it UBVRI} computations
(Sorokina~et~al.~\cite{sbb}).
In far UV, LA4 looks out not so hot, yet it can be detected in far UV earlier
than in visual light.

One should be cautious applying our results directly to observations in UV
range.
A large fraction of SNe~Ia shows a significant correlation with star-forming
regions (Bartunov~et~al. \cite{btsf}; McMillan \& Ciardullo~\cite{mcmici};
Bartunov \& Tsvetkov \cite{batsv}).
The circumstellar medium in those regions can absorb radiation, especially in
UV band.
In this case our predictions should be used as an input for calculations of
reprocessing of UV photons to redder wavelengths.

Certainly, more thorough investigation of the UV emission from SNe~Ia has to
be done.
It is still necessary to calculate UV light curves
of wider range of SNe~Ia models and to predict their UV spectra.
This work is worth doing because, as it is seen from this Letter,
near-UV and far-UV observations with modern UV space telescopes,
when combined with standard {\it UBVRI} study, could be used as an efficient
means to distinguish modes of explosion of thermonuclear supernovae
leading us to better understanding of these phenomena.

\begin{acknowledgements}
We would like to thank W.~Hillebrandt, who has encouraged us to do this
work. We are grateful to E.~Livne, K.~Nomoto, P.~Ruiz-Lapuente, and S.~Woosley
for constructing the SNe~Ia models used in our analysis, and we are thankful
to P.~Lundqvist for his comments on the possibilities of modern space
telescopes.
The work of SB is supported in Russia by ISTC grant 370--97, in Germany by
MPA, and in Japan by ILE, Osaka University.
\end{acknowledgements}


\begin{thebibliography}{}

\bibitem[1997]{batsv} Bartunov~O.S., Tsvetkov~D.Yu. 1997, in Thermonuclear
    Supernovae, eds. P.~Ruiz-Lapuente, R.~Canal, J.~Isern (Dordrecht:
        Kluwer), 87

\bibitem[1994]{btsf} Bartunov~O.S., Tsvetkov~D.Yu., Filimonova~I.V. 1994, PASP
    106, 1276

\bibitem[1996]{blinn96} Blinnikov~S.I. 1996,
  Astron. Letters.    22,  79

\bibitem[1997]{blinn97} Blinnikov~S.I. 1997, in Thermonuclear Supernovae, eds.
        P.~Ruiz-Lapuente, R.~Canal, J.~Isern  (Dordrecht:
        Kluwer), 589

\bibitem[1993]{blba} Blinnikov~S.I., Bartunov~O.S. 1993,
   A\&A 273, 106

\bibitem[1998]{bebpw} Blinnikov~S.I., Eastman~R., Bartunov~O.S.,
    Popolitov~V.A., Woosley~S.E. 1998, ApJ 496, 454

\bibitem[1993]{easpi} Eastman~R.G., Pinto~P.A. 1993,
             ApJ 412,  731

\bibitem[1996]{hoekho} H\"oflich~P., Khokhlov~A. 1996, ApJ 457, 500

\bibitem[1992]{jefetal} Jeffery~D.J., Leibundgut~B., Kirshner~R.P., et~al.,
    1992, ApJ 397, 304

\bibitem[1993]{kiretal} Kirshner~R.P., Jeffery~D.J., Leibundgut~B., et~al.,
    1993, ApJ 415, 589

\bibitem[1991]{leietal} Leibundgut~B., Kirshner~R.P., Filippenko~A.V., et~al.,
    1991, ApJ 371, L23

\bibitem[1990]{liv} Livne~E. 1990, ApJ 354, L53

\bibitem[1995]{liar} Livne~E., Arnett~D. 1995, ApJ 452, 62

\bibitem[1991]{ligl} Livne~E., Glasner~A.S. 1991, ApJ 370, 272

\bibitem[1996]{mcmici} McMillan~R.J., Ciardullo~R. 1996, ApJ 473, 707

\bibitem[1984]{nthyo} Nomoto~K., Thielemann~F.--K., Yokoi~K. 1984,
        ApJ 286, 644

\bibitem[1997]{nugetal} Nugent~P., Baron~E., Branch~D., Fisher~A.,
    Hauschildt~P.H., 1997, ApJ 485, 812

\bibitem[1995]{punetal} Pun~C.S.J., Kirshner~R.P., Sonneborn~G., et al., 1995,
ApJS 99, 223

\bibitem[1993]{rula91bg} Ruiz-Lapuente~P., Jeffery~D.J., Challis~P.M.,
et al., 1993, Nature 365, 728

\bibitem[1995]{rulaetal} Ruiz-Lapuente~P., Kirshner~R.P., Phillips~M.M.,
    et al., 1995, ApJ 439, 60

\bibitem[1999]{rulaproc}  Ruiz-Lapuente~P., Blinnikov~S., Canal~R., et al.,
    1999, in Proc.
      Cosmology with type Ia Supernovae, eds. J.~Niemeyer \&
J.~Truran, in press

\bibitem[1999]{sem} Sembach~K.R. 1999, in ASP Conference Series 166, Stromlo
    Workshop on High-Velocity Clouds, eds. B.K.~Gibson \& M.E.~Putman, 243

\bibitem[2000]{sbb} Sorokina~E.I., Blinnikov~S.I., Bartunov~O.S. 2000,
    Astron. Letters 26, 67 (astro--ph/9906494)

\bibitem[1994a]{wwlowm} Woosley~S.E., Weaver~T.A. 1994a, ApJ 423, 371

\bibitem[1994b]{wwmods} Woosley~S.E., Weaver~T.A. 1994b, in Supernovae,
        eds. J.~Audouze et al. (Amsterdam: Elsevier),
        63

\end{thebibliography}
\end{document}